\documentstyle[twoside,fleqn,espcrc2]{article}
\include{epsf}
\title{Hopping expansion as a tool for handling dual variables in
lattice models}

\author{Christof Gattringer 
\address{Massachusetts Institute of Technology, 
Center for Theoretical Physics \\
77 Massachusetts Avenue, Cambridge  MA 02139 USA}
\thanks{This work was partly supported by Fonds zur F\"orderung der 
wissenschaftlichen Forschung, project J1577-PHY.}}

\begin{document}

\begin{abstract}
The hopping expansion of 8-vertex models in their Grassmann 
representation is studied. We use the functional
similarity of the Ising model in this expansion
with the hopping expansion of 2-D 
Wilson fermions to show that the lattice fermions are equivalent to 
the Self-avoiding Loop Model at bending rigidity $1/\sqrt{2}$. 
\end{abstract}

\maketitle
\section{Grasssmann representation for 8-vertex models and the hopping 
expansion}
\begin{figure}[htp]
\vspace*{-5mm}
\centerline{\hspace*{-2mm}
\epsfysize=3.8cm \epsfbox[ 0 0 303 228 ] {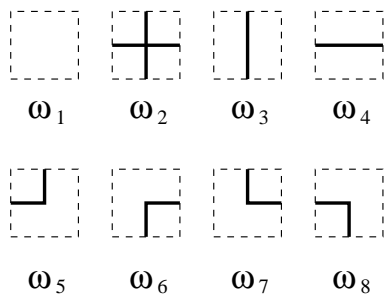}}
\vspace*{-5mm}
\caption{ {\sl Vertices and their weights in the 8-vertex model.}
\label{tiles}}
\end{figure}

Many lattice models have representations in terms of dual variables.
E.g.~the partition function of the 2-dimensional Ising model can be 
written as a sum over all possible Peierls contours with a weight 
given by $e^{-2\beta}$ to the power of the length of the contours.
A generalized model which includes many dual models of lattice systems 
is the 8-vertex model \cite{FaWu69,Ba82}. In the 8-vertex model
the dual variables (loops) are 
decomposed into their elements and
each of those elements (vertices) is assigned a weight. The vertices
can be viewed as 8 different tiles and are shown in Fig.~\ref{tiles}. 
The figure also gives the weights $\omega_i$ in the notation of 
\cite{FaWu69}.
\\

The loops are obtained as tilings of the plane such that there is never 
an open end for the fat line. The partition 
function of the 8-vertex model is then given by 
\begin{equation}
Z_{8v} \; = \; \sum_{loops} \; \prod_{i = 1}^8 \; ( \omega_i)^{n_i} \; ,
\label{z8v}
\end{equation}
where $\omega_i$ is the weight of the i-th vertex, and $n_i$ gives the
abundance of this vertex in the loop configuration.

It is known \cite{Sa80,FrSrSu80}, that the 8-vertex model has a 
representation as an integral over Grassmann variables
\begin{equation}
Z_{8v} \; = \; \int [ d \eta ] \; e^{-S[\eta]} \; .
\label{grassint}
\end{equation}
Here to each site $x$ of the dual lattice a vector of 4 Grassmann variables
\begin{equation}
\eta(x) \; = \; (\eta_{+1}(x), \eta_{-1}(x), \eta_{+2}(x), \eta_{-2}(x))^T
\; , \label{grassvect}
\end{equation}
was assigned. The action $S$ is a sum of pro\-pa\-ga\-tor-, corner- and
monomer-terms ($S = S_p + S_c + S_m$), with
\begin{eqnarray}
S_p = \sum_x \!\!\! & \!\!\! [ \!\!\! & \!\!\!
a \; \eta_{+1}(x) \eta_{-1}(x + \hat{1}) 
\nonumber \\
& \!\!\! + \!\!\! & \!\!\!
b \; \eta_{+2}(x) \eta_{-2}(x + \hat{2}) ] \; , 
\nonumber \\
S_c = \sum_x \!\!\! & \!\!\! [ \!\!\! & \!\!\!
c \; \eta_{+1}(x) \eta_{-2}(x) + d \; \eta_{+2}(x) \eta_{-1}(x)
\nonumber \\ \!\!\! & \!\!\! + \!\!\! & \!\!\!
e \; \eta_{-2}(x) \eta_{-1}(x) + f \; \eta_{+2}(x) \eta_{+1}(x) ] \; ,
\nonumber \\
S_m = \sum_x \!\!\! & \!\!\! [ \!\!\! & \!\!\!g \; 
\eta_{-1}(x) \eta_{+1}(x) + h \; 
\eta_{-2}(x) \eta_{+2}(x) ].
\label{grassact}
\end{eqnarray}
When expressing the weights $\omega_i$ through the coefficients 
$a,b, ... \; h$ as 
\[
\omega_1 = - cd - ef + gh \; , \; \omega_2 = - ab \; , \; 
\omega_3 = bg \; , \] \[ \omega_4 = ah \; , \; 
\omega_5 = c \sqrt{ab} \; , \; \omega_6 = d \sqrt{ab} \; , \]
\begin{equation}
 \omega_7 = e \sqrt{ab} \; , 
\; \omega_8 = f \sqrt{ab} \; , 
\label{parameters}
\end{equation}
it can be shown by explicit expansion that the Grassmann integral
(\ref{grassint}) reproduces the partition function 
(\ref{z8v}). Since (\ref{grassact}) can be diagonalized by Fourier
transform, finding a Grassmann representation (\ref{grassint}), 
(\ref{grassact}) corresponds to solving the model. This is reflected by the 
fact, that the choice (\ref{parameters}) obeys the free fermion condition
$\omega_1 \omega_2 + \omega_3 \omega_4 = \omega_5 \omega_6 + 
\omega_7 \omega_8$, which is a sufficient condition for an analytic 
solution of the 8-vertex model \cite{FaWu69}.

Anti-symmetrization turns the action into a quadratic form
$S = \frac{1}{2} \sum_{x,y} \eta(x)^T K(x,y) \eta(y)$ with kernel
\begin{equation} 
K \; = \; M \; + \; \sum_{\mu = \pm 1}^{\pm 2} P_\mu \; ,
\end{equation}
where $M, P_\mu$ have lattice indices $x,y$ and indices $i,j$ acting
on the Grassmann 4-vectors (\ref{grassvect}). These matrices are given by
\begin{equation}
 M (x,y) \; = \; \delta_{x,y} \; 
\left( \begin{array}{cccc} 
 0 & -g & -f & +c \\ 
+g &  0 & -d & -e \\ 
+f & +d &  0 & -h \\ 
-c & +e & +h & 0 
\end{array} \right) \; ,
\end{equation}
and
\[
(P_{+1}(x,y))_{i,j} \; = \; a \; 
\delta_{x+\hat{1},y} \delta_{i, 1} \delta_{j,2} \; , 
\]
\[ 
(P_{+2}(x,y))_{i,j} \;  = \; b \; 
\delta_{x+\hat{2},y} \delta_{i, 3} \delta_{j,4} \; , 
\]
\begin{equation} 
P_{-1} = - P_{+1}^T \; \; \; , \; \; \; P_{-2} = - P_{+2}^T \; .
\end{equation}
Besides the termwise expansion which gives the partition function 
(\ref{z8v}), the Grassmann integral (\ref{grassint}) can also be 
evaluated as a Pfaffian giving
\begin{equation}
Z_{8v} \; = \; \mbox{Pf} \; K \; = \; \sqrt{ \det K } \; .
\end{equation} 
In the second equality we used the fact that $K$ is anti-symmetric and thus
its Pfaffian reduces to the root of a determinant. The matrix $M$ has 
determinant $(gh - ef - cd)^2$ which after inserting (\ref{parameters}) 
and using the free fermion condition reduces to $\det M = \omega_1^2$.
Thus for 8-vertex models with $\omega_1 \neq 0$, the matrix $M$ can be 
inverted and one can transform to
\[
Z_{8v} = \sqrt{\det{M}} \sqrt{\det[ 1 + H ]} \;  = \]
\begin{equation} 
\exp \left(
- \frac{1}{2} \sum_{n=1}^{\infty} \frac{(-1)^n}{n} \; \mbox{Tr} 
H^n \right) \; ,
\label{hopexp}
\end{equation}
where we used the well known formula for the determinant as the
exponential of the trace of the logarithm and defined the hopping matrix
$H$ as
\begin{equation}
H \; \equiv \; \sum_{\mu = \pm 1}^{\pm 2} M^{-1} P_\mu \; .
\label{hopmat}
\end{equation}
The series in the exponent converges for values of the parameters where
$\parallel \! H \! \parallel < 1$. Since the matrices $P_\mu$ contain the 
hopping factors $\delta_{x+\mu,y}$ only terms that correspond to closed
loops can contribute to (\ref{hopexp}). These closed loops are however 
related to the loops used in the original definition (\ref{z8v}) of the
model and thus the exponent in (\ref{hopexp}) is in principle the free 
energy of the model expressed in terms of loops. The only remaining 
problem is to explicitly solve the traces over powers of the hopping
matrix $H$ in (\ref{hopexp}). For the general case this will be discussed 
elsewhere \cite{Ga98}. 

The Ising model can be obtained from the 8-vertex model by setting
\begin{equation}
\omega_1 = 1 \; , \; \omega_2 = e^{-4\beta} \; , \; \omega_j = e^{-2\beta}
\; , \; j \; = 3, ... 8.
\end{equation}
For this case the hopping expansion (\ref{hopexp}) gives \cite{GaJaSe98}
\begin{equation}
Z_i = \exp \left(\sum_l \frac{(-1)^{n(l)}}{I(l)} 
e^{-2\beta |l|} \right) \; .
\label{isingresult}
\end{equation}
Here $l$ runs over all closed, simply connected loops on the
lattice, $n(l)$ denotes the number of self-intersections, $I(l)$
gives the number of times the loop $l$ iterates its entire trace and
$|l|$ is the length of the loop.

\section{Equivalence of 2-D Wilson fermions with the Self-avoiding Loop Model
at bending rigidity $1/\sqrt{2}$}
In this section we give an example how representations of the
type (\ref{hopexp}) can be used. We will show that 2-dimensional Wilson 
fermions on the lattice are equivalent to the Self-avoiding Loop Model at 
bending rigidity $1/\sqrt{2}$. For an alternative derivation of this result 
see Scharnhorst's paper \cite{Sch96}.

The action for 2-dimensional Wilson fermions \cite{Wi74} is given by
\begin{equation}
S \; = \; \overline{\psi} M \psi 
\; \; \; \; \; \; ,  \; \; \; \; \; \; 
M \; = \; 1 \; - \; \kappa Q \; ,
\end{equation}
with 
\begin{equation}
Q(x,y) \; = \; \sum_{\mu = \pm 1}^{\pm 2}
\Gamma_\mu \; \delta_{x+\mu, y} \; ,
\end{equation}
where $\Gamma_{\pm \mu} = \frac{1}{2} [ 1 \mp \sigma_\mu ], 
\; \mu = 1,2$. Here $\sigma_1$, $\sigma_2$ denote the Pauli matrices and the
hopping parameter $\kappa$ is related to the bare fermion 
mass $m$ via $\kappa = (m + 2)^{-1}$. Since the action for the lattice 
fermions is a bilinear form the partition function is not a 
Pfaffian but a determinant
\[
Z_w(\kappa) \; = \; \int [ d \overline{\psi} ] [ d \psi ] \; e^{-S} \; = \; 
\det M \; = \]
\begin{equation}
\det[1 - \kappa Q] \; = \; 
\exp \left( - \sum_{n=1}^{\infty} \frac{\kappa^n}{n} \mbox{Tr} Q^n 
\right)\; .
\end{equation}
Again the contributions to the traces in the exponent correspond to closed
loops, and the remaining problem is to compute the traces of the matrices
$\Gamma_\mu$ ordered along the links of the loops. This problem was solved 
in \cite{St81} by realizing that the Pauli matrices give rise to a 
representation of rotations on the lattice. The resulting hopping expansion 
is
\begin{equation}
Z_w \; = \; \exp \left( 2  \sum_{l} \kappa^{|l|}
\frac{(-1)^{n(l)}}{I(l)} \left(\frac{1}{\sqrt2} \right)^{c(l)}  
\right) \; .
\label{wilsonhopp} 
\end{equation}
This result is very similar to Formula (\ref{isingresult}) 
for the Ising model,
with the only difference of an extra weight $1/\sqrt{2}$ for each corner of
a loop $l$ (here $c(l)$ denotes the number of corners in a loop $l$).  

When expanding the exponential functions in (\ref{isingresult}) or
(\ref{wilsonhopp}) one obtains contributions from products of  
loops $l$ appearing in the exponent, such that in these products 
some links are multiply occupied. In addition already single loops 
$l$ can retrace some of their links such that they are multiply 
occupied. For the Ising model we know that all the contributions with
multiply occupied links cancel since the Peierls contour representation
discussed in the beginning allows only for simply occupied links (see 
\cite{GaJaSe98} for a detailed discussion). This cancellation mechanism 
is independent of corner weights since two loop configurations which
differ in their number of corners necessarily have different occupancy 
numbers
at some links and thus only contributions with equal numbers of corners
can cancel each other. Hence, using the similarity between (\ref{wilsonhopp})
and (\ref{isingresult}) we can conclude that also in an expansion of the 
exponential in (\ref{wilsonhopp}), terms with multiply occupied links
cancel. The remaining terms inherit the corner factors $1/\sqrt{2}$, and the
self-intersection factor $(-1)^{n(l)}$ in (\ref{wilsonhopp}) leads to a
cancellation of terms with self-intersections (i.e.~the corresponding 
vertex model has weight $\omega_2 = 0$ and thus the loops are self-avoiding).
Thus we end up with (see \cite{Ga98} for the detailed proof)
\begin{equation}
\sqrt{Z_{w} (\kappa)} \; = \;  Z_{salm} \left(\kappa, \frac{1}{\sqrt{2}}
\right) \; , 
\end{equation} 
with the Self-avoiding Loop model defined as 
\begin{equation}
Z_{salm}(z, \eta) \; = \; \sum_{\gamma} 
z^{|\gamma|} \; \eta^{c(\gamma)} \; .
\label{salmdef}
\end{equation}
Here $\gamma$ runs over all self-avoiding contours, $c(\gamma)$ gives the 
number of corners, $\eta$ is the bending rigidity, $z$ is the monomer weight 
and $|\gamma|$ denotes
the length of the contour.

\end{document}